\documentclass[prl,twocolumn,showpacs,groupedaddress,10pt,nofootinbib]{revtex4-1} 
\usepackage{amsfonts,amsmath,amssymb,graphicx}

\usepackage{color,times}
\usepackage{multibib}

\definecolor{darkblue}{rgb}{0, 0, 0.8}
\usepackage[colorlinks=true, breaklinks=true, linkcolor=darkblue, citecolor=darkblue, urlcolor=darkblue]{hyperref}
\newcommand{\ket}[1]{\left| #1\right\rangle}
\newcommand{\bra}[1]{\left\langle #1\right|}
\newcommand{\bs}{\boldsymbol}
\newcommand{\udd}{\ket{\uparrow\downarrow\downarrow}}
\newcommand{\dud}{\ket{\downarrow\uparrow\downarrow}}
\newcommand{\ddu}{\ket{\downarrow\downarrow\uparrow}}

\begin{document}

\title{Coherent Excitation Transfer in a Spin Chain of Three Rydberg Atoms}

\author{Daniel Barredo}
\author{Henning Labuhn}
\author{Sylvain Ravets}
\author{Thierry Lahaye}
\author{Antoine Browaeys}

\affiliation{Laboratoire Charles Fabry, UMR 8501, Institut d'Optique, CNRS, Universit\'e Paris Sud 11,\\
2 avenue Augustin Fresnel, 91127 Palaiseau cedex, France }

\author{Charles S. Adams}

\affiliation{
Joint Quantum Centre (JQC) Durham-Newcastle, Department of Physics, Durham University, Durham, DH1 3LE, United Kingdom }

\begin{abstract}
We study coherent excitation hopping in a spin chain realized using highly excited individually addressable Rydberg atoms. The dynamics are fully described in terms of an \textit{XY} spin Hamiltonian with a long range resonant dipole-dipole coupling that scales as the inverse third power of the lattice spacing, $C_3/R^3$. The experimental data demonstrate the importance of next neighbor interactions which are manifest as revivals in the excitation dynamics. The results suggest that arrays of Rydberg atoms are ideally suited to large scale, high-fidelity quantum simulation of spin dynamics.
\end{abstract}

\maketitle 


Spin Hamiltonians, introduced in the early days of quantum mechanics to explain ferromagnetism, are widely used to study quantum magnetism~\cite{auerbach1994}. Assemblies of interacting, localized spins are a paradigm of quantum many-body systems, where the interplay between interactions and geometry-induced frustration creates a wealth of intriguing quantum phases. Many other phenomena, such as coherent energy transfer, photochemistry or photosynthesis~\cite{collini2013}, can also be described using spin Hamiltonians. However, despite this fundamental significance, exact analytical solutions are known only for the simplest cases, and numerical simulations of strongly correlated spin systems are notoriously difficult.

For those reasons, quantum simulation of spin Hamiltonians by controllable systems raises great interest. Recently, various approaches were followed to simulate spin systems using tools of atomic physics~\cite{lewenstein2012}, such as cold atoms~\cite{simon2011,fukuhara2013,depaz2013} or polar molecules~\cite{yan2013} in optical lattices, interacting via weak exchange or dipole-dipole interactions, or trapped ions with engineered effective interactions~\cite{kim2010, richerme2014, jurcevic2014}. As compared to their condensed-matter counterparts, the spin couplings can be long range, which gives rise to new properties~\cite{hauke2010,peter2012,hazzard2014,avellino2006}.

Rydberg atoms are a promising alternative platform for quantum simulation~\cite{saffman2010,weimer2010}. In particular, they allow implementing various spin-$1/2$ Hamiltonians on two-dimensional lattices with strong couplings, in the MHz range~\cite{barredo2014,schauss2014}. Rydberg systems interacting through van der Waals interactions can be described by Ising-type Hamiltonians $H=\sum_{ij}V_{ij}\sigma_i^z \sigma_j^z$  where $\sigma^z$ is the $z$-Pauli matrix acting in the (pseudo-) spin Hilbert space, and $V_{ij}\sim|{\bs r}_i-{\bs r}_j|^{-6}$, where ${\bs r}_{i}$ denotes the position of atom $i$~\cite{weimar2008,pohl2010,lesanovsky2011,barredo2014,schauss2014}. On the other hand, spin-exchange, or \textit{XY}, spin Hamiltonians of the form $H=\sum_{ij}V_{ij}(\sigma_i^+ \sigma_j^- + \sigma_i^- \sigma_j^+)$, where $\sigma^\pm=\sigma^x\pm i\sigma^y$ are spin-flip operators and $V_{ij}\sim|{\bs r}_i-{\bs r}_j|^{-3}$, can be realized by using two different Rydberg states, interacting directly via the resonant dipole-dipole interaction. However in this case, only incoherent transfer of excitations has been observed so far, due to the random atomic positions in the ensembles used in experiments~\cite{mourachko1998,anderson1998,heuvell2008,betelli2013,guenter2013,maxwell2013}.

In this Letter, we study the coherent dynamics of a spin excitation in a chain of three Rydberg atoms. The dipole-dipole interaction between atoms is given by the \textit{XY} Hamiltonian~\cite{Supplementary}
\begin{equation}
H = \frac{1}{2} \sum_{i \neq j} \frac{C_3}{R_{ij}^3} \left(\sigma_i^+ \sigma_j^- + \sigma_i^- \sigma_j^+\right),
\label{eq:dd}
\end{equation}
where $R_{ij}=|{\bs r}_i-{\bs r}_j|$ is the distance between atoms $i$ and~$j$. 
We calibrate the spin-spin coupling between two Rydberg atoms by investigating the temporal evolution of two Rydberg atoms prepared in the state $\ket{\uparrow\downarrow }$, as a function of distance $R$ between the atoms, up to $R\simeq50\;\mu$m. We then use three Rydberg atoms prepared in $\udd$ and study the propagation of the excitation through this minimalistic spin chain, observing the effect of long-range hopping of the excitation. The agreement between experimental data and the \textit{XY} model without adjustable parameters validates our setup as a future quantum simulator for systems of many spins in arbitrary two-dimensional arrays~\nocite{Tuchendler2008}.

\begin{figure}[t]
\centering
\includegraphics[width=\linewidth]{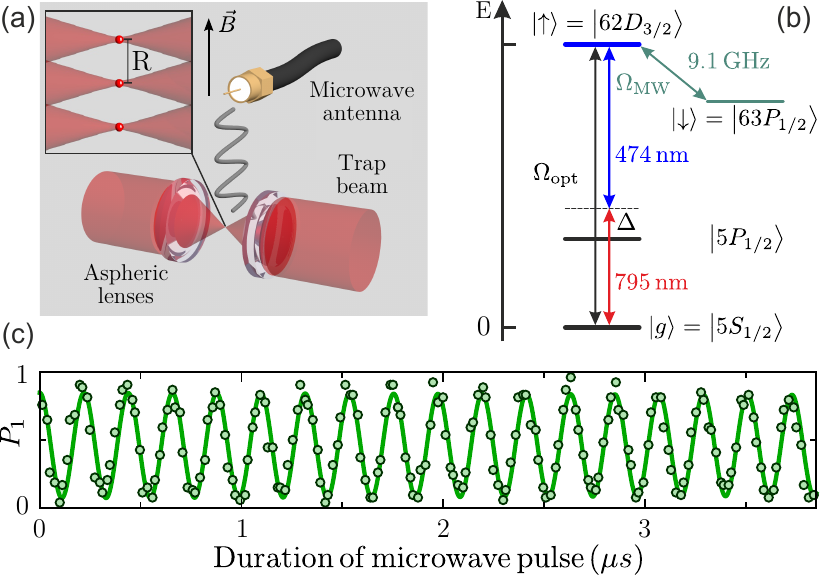}
\caption{(color online). (a)~Individual $^{87}{\rm Rb}$ atoms in microtraps aligned along the quantization axis, defined by a $B=6$~G magnetic field. (b)~Excitation lasers couple the ground state $\ket{g} = \ket{5 S_{1/2},F=2,m_F=2}$ and the Rydberg state $\ket{\uparrow}=\ket{62 D_{3/2},m_j=3/2}$ with an effective Rabi frequency $\Omega_{\rm opt}$. Microwaves couple $\ket{\uparrow}$ to $\ket{\downarrow} = \ket{63 P_{1/2},m_J=1/2}$, with Rabi frequency $\Omega_{\rm MW}$.  (c)~Microwave-driven Rabi oscillation of a single atom  between $\ket{\uparrow}$ and $\ket{\downarrow}$, yielding $\Omega_{\rm MW} = 2 \pi \times 4.6$~MHz. }
\label{fig:fig1}
\end{figure}

\begin{figure}[t]
\centering
\includegraphics[width=0.9\linewidth]{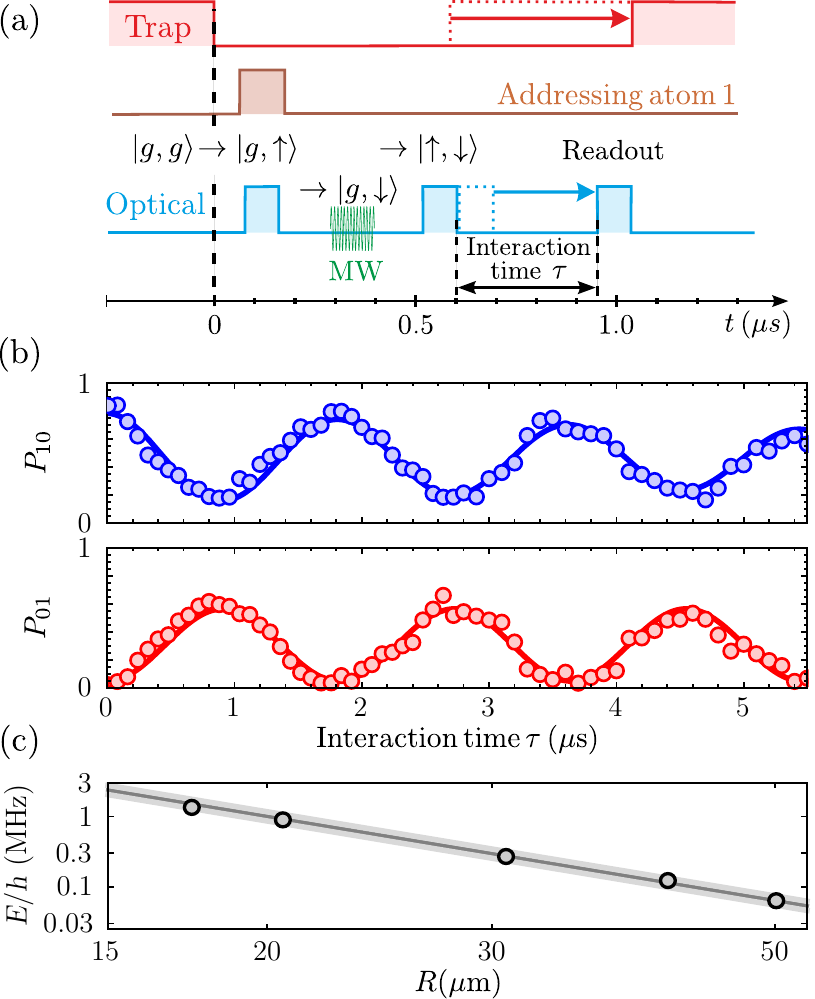}
\caption{(color online). 
(a) Sequence to observe spin exchange between two atoms. (b) Excitation hopping between states $\ket{\uparrow \downarrow}$ (blue disks) and $\ket{\downarrow \uparrow}$ (red disks) of two atoms separated by $R=30\; \mu$m. Solid lines are sinusoidal fits, with frequency $2E/h$. (c) Interaction energy $E$ (circles) versus $R$. Error bars are smaller than the symbols size. The line shows the theoretical prediction $C_3/R^3$ with $C_3^{\rm th}=7965\;{\rm MHz}\; \mu {\rm m}^3$. The shaded area corresponds to our systematic 5\% uncertainty in the calibration of~$R$. }
\label{fig:fig2}
\end{figure}

The experimental setup, shown in Fig.~\ref{fig:fig1}(a), is detailed in Ref.~\cite{beguin2013}. Briefly, we focus a red-detuned trapping beam with an aspheric lens into a magneto-optical trap of $^{87}$Rb, to a waist $\simeq 1\;\mu$m.  Multiple traps at arbitrary distances are created by imprinting an appropriate phase on the trapping beam with a spatial light modulator~\cite{nogrette2014}. Beccause of fast light-assisted collisions in the small trapping volume, at most one atom is present in each trap. The temperature of the trapped atoms is approximately 50~$\mu$K. A 6~G magnetic field  defines the quantization axis~\cite{Note1}.

\begin{figure*}
\centering
\includegraphics[width=\linewidth]{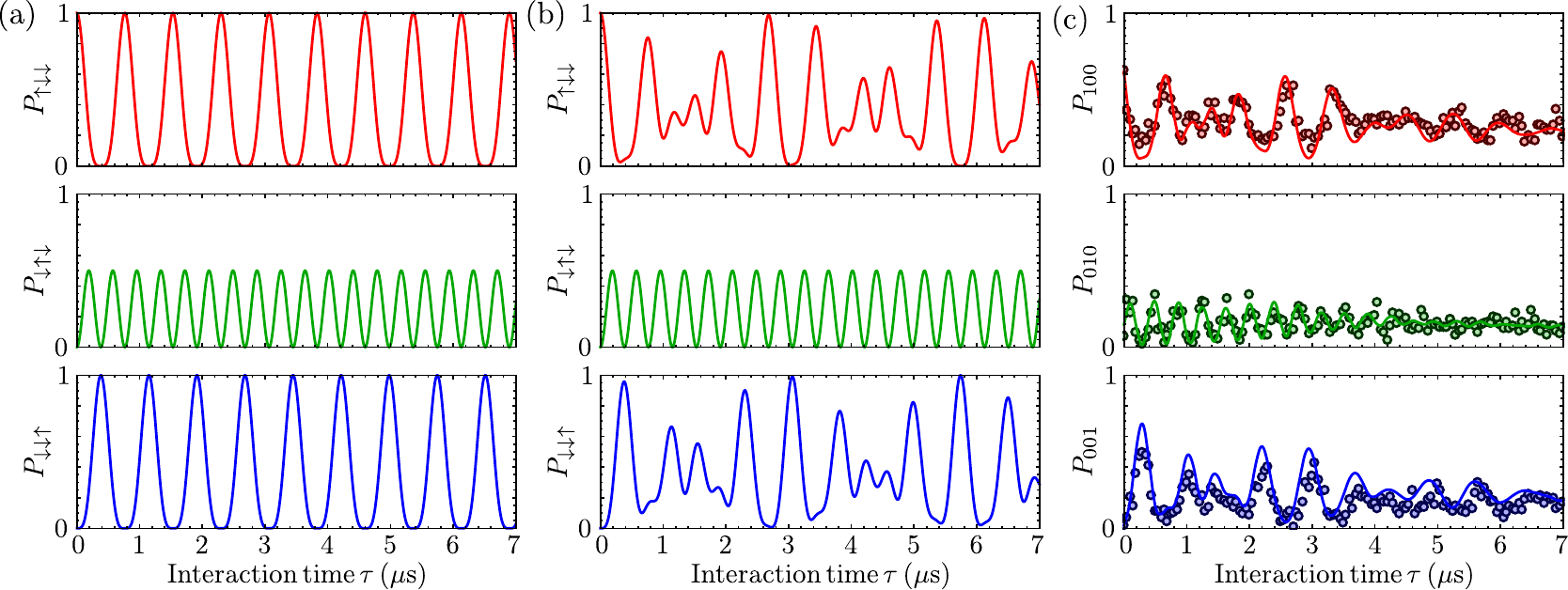}
\caption{(color online). Spin excitation transfer along a chain of three Rydberg atoms with nearest-neighbor separation of 20~$\mu$m. (a) Theoretical dynamics for a system initially prepared in $\ket{\uparrow \downarrow \downarrow}$, and evolving under a Hamiltonian similar to (\ref{eq:dd}), but with only nearest-neighbor interactions. (b) The same as (a), but for the the full Hamiltonian (\ref{eq:dd}), including long-range interactions. (c) Experimental data (points) and prediction of the model taking into account experimental imperfections (see text), with no adjustable parameters. For perfect preparation and readout, the probabilities $P_{\uparrow\downarrow\downarrow}$ (respectively $P_{\downarrow\uparrow\downarrow}$, $P_{\downarrow\downarrow\uparrow}$) and $P_{100}$ (respectively $P_{010}$, $P_{001}$) would coincide.} 
\label{fig:fig3}
\end{figure*}

We encode the two spin states in the Rydberg states $\ket{\uparrow} = \ket{62D_{3/2},m_j=3/2}$ and $\ket{\downarrow} = \ket{63P_{1/2},m_j=1/2}$ [see Fig. \ref{fig:fig1}(b)]. We trigger an experiment when an atom is detected in each trap. To prepare the atoms in a desired spin state, we first optically pump them in $\ket{g}=\ket{5S_{1/2},F=2,m_F=2}$. We then switch off the traps to avoid inhomogeneous light shifts, and excite the atoms to $\ket{\uparrow} = \ket{62D_{3/2},m_J=3/2}$ via a two-photon transition (wavelengths 795 and 474~nm, with polarizations $\pi$ and $\sigma^+$, respectively), detuned from the intermediate state $\ket{5P_{1/2},F=2,m_F=2}$ by $\Delta \simeq 2 \pi \times 740$~MHz. From the $\ket{\uparrow}$ state the atom can be transferred to $\ket{\downarrow} = \ket{63P_{1/2},m_J=1/2}$ using resonant microwaves at $\simeq 9.131$~GHz, emitted by an antenna outside the vacuum chamber. 

To read out the state of an atom at the end of a sequence, we switch on the excitation lasers, coupling only $\ket{\uparrow}$ back to the ground state. We then turn on the dipole traps to recapture ground-state atoms, while atoms in Rydberg states remain untrapped, and detect atoms in $\ket{g}$ by fluorescence. Therefore if we detect an atom in its trap at the end of a sequence, we assume it was in $\ket{\uparrow}$, while a loss corresponds to the $\ket{\downarrow}$ state. We reconstruct all the $2^N$ probabilities $P_{i_1\ldots i_k\ldots i_N}$ of having $i_k$ atom in trap $k$, with $i_k=0$ or 1, for our $N$-trap system (with $N=1, 2$, or $3$) by repeating the experiment typically 100 times. For instance for $N=3$, $P_{100}$ is the probability to recapture an atom in trap 1, while recapturing none in traps 2 and 3. The statistical error on the determination of the probabilities is below $5\%$. Figure~\ref{fig:fig1}(c) illustrates the coherent spin manipulation for a single atom, by showing Rabi oscillations between $\ket{\uparrow}$ and $\ket{\downarrow}$: the probability $P_1$ to recapture the atom oscillates with a frequency $\Omega_{\rm MW} \simeq 2\pi \times 4.6$~MHz. In $4\;\mu{\rm s}$, we induce more than 35 spin flips without observing noticeable damping. 

We first use two atoms, aligned along the quantization axis, to directly measure the coupling between two spins as a function of their distance. The sequence  is shown in Fig. \ref{fig:fig2}(a). We illuminate atom 1 with an addressing beam~\cite{labuhn2014} which induces a $20$~MHz light shift, making it off resonant to the global Rydberg excitation. Atom 2 is excited to $\ket{\uparrow}$, and then transferred to $\ket{\downarrow}$ using microwaves. Subsequently, atom 1 is optically excited to the $\ket{\uparrow}$ state with the addressing beam  switched off (atom 2 in  $\ket{\downarrow}$ is not affected by the Rydberg excitation pulse). We let the system evolve for an adjustable time $\tau$ and read out the final state by deexciting $\ket{\uparrow}$ back to $\ket{g}$. In the absence of experimental imperfections (see~\cite{Supplementary}, section S.4), $P_{10}$ (respectively $P_{01}$) would give the population of $\ket{\uparrow\downarrow}$ (respectively $\ket{\downarrow\uparrow}$).

The evolution of $P_{10}(\tau)$ and $P_{01}(\tau)$ for two atoms prepared in $\ket{\uparrow \downarrow}$ separated by 30~$\mu$m is shown in Fig.~\ref{fig:fig2}(b). The spin excitation oscillates back and forth between the two atoms, with a frequency $2E/h \approx 0.52$~MHz. The finite contrast is essentially due to spontaneous emission via the intermediate $\ket{5P_{1/2}}$ state during preparation and readout, which limits the oscillation amplitude to about $60\%$, and, to a lesser extent, to the onset of dipolar interactions during the second excitation pulse \cite{Supplementary}. We then repeat the same experiment for several values of the distance $R$ between the atoms, and  observe spin-exchange oscillations for distances as large as 50~$\mu$m. Figure~\ref{fig:fig2}(c) shows the measured interaction energies as a function of $R$, together with the expected $C_3/R^3$ behavior (solid line) for the theoretical value $C_3^{\rm th} = 7965\;{\rm MHz}\; \mu {\rm m}^3$ of the $C_3$ coefficient, calculated from the dipole matrix elements $\bra{\uparrow}\hat{d}_{\pm 1}\ket{\downarrow}$ \cite{beguin2013,reinhard2007}. A power-law fit to the data (not shown) gives an exponent $-2.93\pm0.20$. Fixing the exponent to $-3$ gives $C_3^{\rm exp}=7950\pm 130 \;{\rm MHz}\; \mu {\rm m}^3$. The agreement between data and theory is excellent.

We now extend the system to a three-spin chain, with a distance $R=20\;\mu$m between the atoms. The sequence is similar to that in Fig.~\ref{fig:fig2}(a) for two atoms, except that we now use microwave transfer for atoms 2 and 3 to prepare $\ket{g \downarrow \downarrow}$. Here, the van der Waals interaction between the two atoms in $\ket{\uparrow}$ is only $\sim 10$~kHz for $R=20\;{\rm \mu m}$, and thus no blockade effect arises during excitation. We then excite atom~1 to prepare $\ket{\uparrow \downarrow \downarrow}$.

We first analyze theoretically the evolution of the system. Assuming that the initial state is $\ket{\psi(0)} = \udd$, the dynamics induced by the \textit{XY} Hamiltonian (\ref{eq:dd}), which conserves the total magnetization $\sum_i\sigma^z_i$, occurs within the subspace spanned by $\{\udd,\dud,\ddu\}$. Figures~\ref{fig:fig3}(a) and (b) show the calculated dynamics of the spin excitation, which moves back and forth between the extreme sites. Figure~\ref{fig:fig3}(a) corresponds to the case where only nearest-neighbor interactions are retained in (\ref{eq:dd}). Periodic, fully contrasted oscillations at a frequency $\sqrt{2}C_3/R^3$ are expected for the population of the extreme sites, while the population of $\dud$ oscillates twice as fast between $0$ and $1/2$. In contrast, in Fig.\ref{fig:fig3}(b), the full Hamiltonian (\ref{eq:dd}) is simulated, including the interaction between extreme sites. One observes a clear signature of this long-range coupling, as the dynamics now becomes aperiodic for the populations of $\udd$ and $\ddu$. The interplay of the couplings $C_3/R^3$ and $C_3/(8R^3)$ between nearest- and next-nearest neighbors makes the eigenvalues of (\ref{eq:dd}) incommensurate. The back-and-forth exchange of excitation is thus modulated by a slowly varying envelope due to the beating of these frequencies. 

Figure~\ref{fig:fig3}(c) shows the experimental results for $P_{100}$, $P_{010}$, and $P_{001}$ (symbols). We observe qualitative agreement with Fig.~\ref{fig:fig3}(b), in particular the ``collapse and revival'' in the dynamics showing the effects of the long-range coupling. However, one notices differences with the ideal case: (i) the preparation is imperfect, as one starts with a significant population in $\dud$, (ii) this, together with imperfect readout~\cite{Supplementary}, reduces the overall amplitude of the oscillations, and (iii) the oscillations show some damping, which becomes significant for $\tau\geq 4\;\mu{\rm s}$.  

Imperfect preparation and readout stem from the fact that, in addition to the spontaneous emission via the intermediate state during the optical pulses, the Rabi frequency for  optical excitation ($\simeq 5.3$~MHz) of atom 1 from $\ket{g}$ to $\ket{\uparrow}$ is not much higher than the interaction ($\simeq0.92\;{\rm MHz}$ for $R=20\;{\rm\mu m}$). Thus, during the excitation of atom 1, the spin excitation already has a significant probability to hop to atom 2. The damping essentially arises from the finite temperature of the atoms, which leads to changes in the interatomic distances, and thus in the couplings.  

To go beyond this qualitative understanding of the limitations of our ``quantum simulator," we add all known experimental imperfections to the \textit{XY} model \cite{Supplementary}. The result, shown by  solid lines on Fig.~\ref{fig:fig3}(c) accurately reproduces the data with no free parameters. To obtain these curves, we simulate the full sequence, i.e., all three optical (de-)excitation pulses with or without the addressing beam, the microwave pulse, and evolution time, by solving the optical Bloch equations describing the dynamics of the internal states of the atoms, restricted to three states: $\ket{g}$, $\ket{\uparrow}$, and $\ket{\downarrow}$. Dissipation comes from both off resonant excitation of the intermediate $\ket{5P_{1/2}}$ state during the optical excitation pulse, and from the finite lifetimes of the Rydberg states (101 and 135 $\mu$s for $\ket{\uparrow}$ and $\ket{\downarrow}$, respectively~\cite{beterov2009}). The former effect is treated as an effective damping of the $\ket{g}\leftrightarrow\ket{\uparrow}$ transition, present only during the optical pulses, and with a damping rate chosen to match the damping of single-atom Rabi oscillations performed to calibrate the excitation Rabi frequency $\Omega_{\rm opt}$~\cite{barredo2014}.

We then account for the thermal motion  of the atoms. A first consequence of the finite temperature ($T\simeq50\;{\rm \mu K}$) is that at the beginning of the sequence, the atoms have random positions (the transverse rms extension of the thermal motion in each microtrap, of radial frequency 90~kHz, is about 120~nm) and random velocities ($ 70~{\rm nm}/\mu{\rm s}$ rms). During the sequence, the traps are switched off and the atoms are thus in free flight with their initial velocity.  When solving the optical Bloch equations, we thus first draw the initial positions ${\bs r}_i^0$ and velocities ${\bs v}_i^0$ of each atom~$i$ according to a thermal distribution, and use time-dependent dipolar couplings $C_3/|({\bs r}_i^0+{\bs v}_i^0 t)-({\bs r}_j^0+{\bs v}_j^0 t)|^3$ in Eq.~(\ref{eq:dd})~\cite{Note2}. We then average the results over 100 realizations. This yields a dephasing of the oscillations, resulting in a significant contrast reduction at long  times.

A second effect of the temperature is that an atom has a small probability $\varepsilon(t)$ to leave the trap region during the experiment. In this case, we mistakenly infer that it was in a Rydberg state at the end of the sequence. This leads to a small distortion of the measured populations $P_{ijk}$ ($i,j,k=0,1$)~\cite{shen2012}, that we compute from the actual ones as described in~\cite{barredo2014}. We measure $\varepsilon(t)$ (which increases with the duration $t$ of the sequence, from $\sim1\%$ at $t=0$  up to $\sim20\%$ for $t=7\;{\rm \mu s}$) in a calibration experiment, and then use it to calculate the expected populations from the simulated ones \cite{Supplementary}. 

\begin{figure}[t!]
\centering
\includegraphics[width=\linewidth]{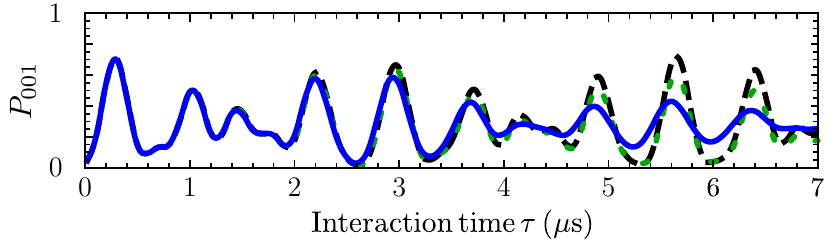}
\caption{(color online). 
Influence of the temperature on $P_{001}(\tau)$: simulated dynamics at zero temperature (black dashed line), and adding either only atom loss (green dotted line), or only atomic motion (blue solid line).
}
\label{fig:fig4}
\end{figure}

Figure \ref{fig:fig4} shows how those two consequences of the finite temperature contribute to the observed damping in the dynamics of $P_{001}$: both have sizable effects, but the dephasing due to fluctuations in the coupling dominates at long times. Reducing the atomic temperature using e.g. Raman cooling~\cite{kaufman2012,thompson2013} would render those effects negligible for our time scales, and allow the realization of a nearly ideal quantum simulator of spin dynamics.

In summary, we have measured the dynamics of a spin excitation in a minimal spin chain of three Rydberg atoms. The evolution of the system is accurately described by an \textit{XY} Hamiltonian without any adjustable parameters. The obtained results are encouraging in view of scaling up the system to a larger number of spins. In particular, the residual motion of the atoms and the level of detection errors would already allow us to observe unambiguously the back-and-forth propagation of an excitation over a chain of $\sim 20 $ atoms~\cite{Supplementary}. However, so far, experiments with more than $\sim 5$ atoms are hampered by the stochastic loading of the traps by single atoms~\cite{nogrette2014}. In future work, we will thus explore various quasideterministic loading schemes that have been demonstrated at the level of a single~\cite{andersen,ebert} or a few~\cite{meschede,birkl} traps. Once this is achieved, our system will allow us to study the equivalent of an assembly of hard-core bosons on a 2D lattice with long-range, anisotropic hopping. We will also study dipolar interactions involving more than only two Rydberg states at an electrically tuned F\"orster resonance \cite{ravets2014}. Our system will be ideal to study exotic phases and frustration in quantum magnetism, excitation hopping in complex networks \cite{wuester2010,mostarda2013} or quantum walks with long-range hopping \cite{cote2006}.

\begin{acknowledgments}
We acknowledge financial support by the EU [ERC Stg Grant ARENA, AQUTE Integrating Project, FET-Open Xtrack Project HAIRS, and EU Marie-Curie Program ITN COHERENCE FP7-PEOPLE-2010-ITN-265031 (H. L.)], and by R\'egion \^Ile-de-France (LUMAT and Triangle de la Physique, projects LAGON and COLISCINA during the stay of C.~S.~A at LCF). C.~S.~A. also acknowledges support from the U.~K. EPSRC and Durham University.
\end{acknowledgments}

\vskip0.7cm

\clearpage

\setcounter{figure}{0}
\setcounter{equation}{0}

 
\renewcommand{\thesection}{S.\arabic{section}}
\renewcommand{\thesubsection}{\thesection.\arabic{subsection}}
 
%
\makeatletter 
\def\tagform@#1{\maketag@@@{(S\ignorespaces#1\unskip\@@italiccorr)}}
\makeatother
 
\makeatletter
\makeatletter \renewcommand{\fnum@figure}
{\figurename~S\thefigure}
\makeatother

\renewcommand{\figurename}{Figure}

\makeatletter
\renewcommand\@cite[1]{[\textcolor{blue}{S}#1]}
\renewcommand\@biblabel[1]{[S#1]}
\makeatother


\onecolumngrid
\noindent
\begin{center}
\large{\bf Supplemental Material: Coherent Excitation Transfer in a Spin Chain of Three Rydberg Atoms \\ }
\end{center}

\section{The dipole-dipole interaction as an \textit{XY} Hamiltonian}

Here we derive, for the sake of completeness, how the dipole-dipole interaction between two Rydberg atoms leads to the \textit{XY} Hamiltonian (1) of the main text. We also emphasize the angular dependence of the $C_3$ coefficient, which is proportional to $1-3\cos^2\theta$.

The interaction between two neutral atoms separated by a distance $R$ (large compared to the atom size, but small compared to the wavelength of the relevant transitions, such that the electrostatic limit applies) can be expressed, to leading order, through the dipole-dipole interaction:
\begin{equation}
\label{Eq:DipDipInt}
\begin{aligned}
V_{\rm{ddi}}=\frac{1}{4 \pi \epsilon_0} \frac{  {\bs{d}}_1 \cdot {\bs{d}}_2 - 3 ({\bs{d}}_1 \cdot {{\bs{n}}}) ({\bs{d}}_2 \cdot {{\bs{n}}})}{ R^3 },
\end{aligned}
\end{equation}
where ${\bs{d}_i} = (d_x,d_y,d_z)$ is the electric dipole moment operator of atom $i$ ($i=1, 2$), and ${\bs{n}}={{\bs{R}}} / R$ is the unit vector connecting the two atoms. We denote the quantization axis by $z$, and the angle between $z$ and ${\bs n}$ by $\theta$. In the spherical basis, it is convenient to use the spherical dipole operators:
\begin{equation}
\label{Eq:DipoleOperatorsDefinition}
\begin{aligned}
\left\{
\begin{array}{rl}
d_{0} &= d_z                                                   \\
d_{+}  &= - \left( d_x+i\,d_y   \right)/\sqrt{2} \\
d_{-}  &=  \left( d_x-i\,d_y   \right)/\sqrt{2} \ .\\
\end{array}
\right.
\end{aligned}
\end{equation}
The operator $d_0$ conserves the magnetic quantum number $m_j$, whereas the operators $d_{\pm}$ change $m_j$ by one ($\Delta m_{j}=\pm 1$). In the spherical basis, the dipole-dipole interaction can be written as:
\begin{eqnarray}
\label{Eq:VdipSph}
V_{\rm{ddi}} = \frac{1}{4 \pi \epsilon_0} \frac{1}{R^3} && \left[ \frac{1-3 \cos ^ 2 \theta}{2} \left(  d_{1,+} d_{2,-} + d_{1,-} d_{2,+} + 2d_{1,0}d_{2,0}  \right) \right. \nonumber \\
&& + \frac{3}{\sqrt{2}} \sin \theta \cos \theta \left( d_{1,+}d_{2,0} - d_{1,-}d_{2,0} +d_{1,0}d_{2,+} - d_{1,0}d_{2,-} \right)\\
&& - \left.  \frac{3}{2} \sin^2 \theta \left( d_{1,+}d_{2,+} + d_{1,-}d_{2,-} \right) \right]. \nonumber \\ \nonumber
\end{eqnarray}
The sum comprises three terms, with different angular dependence, that couple states where the total magnetic quantum number $M= m_j^{(1)}+m_j^{(2)}$ changes, respectively, by $\Delta M=0$, $\Delta M=\pm 1$, and $\Delta M=\pm 2$. 

We now restrict ourselves to only two states $\ket{\uparrow}$ and  $\ket{\downarrow}$ of the form $\ket{n,L,J,m_j}$ and $\ket{n',L',J',m_j'}$ that fulfill the selection rules for the dipole operator $d_q$ ($q=0,\pm1$), i.e. $\Delta L =\pm1$, $ \Delta J=0,\pm1$, and $\Delta m_j=q$ (in the main text, they are $ \ket{62D_{3/2},m_j=3/2}$ and  $\ket{63P_{1/2},m_j=1/2}$). In the two-atom basis $\left\{\ket{\uparrow\uparrow},\ket{\uparrow\downarrow},\ket{\downarrow\uparrow},\ket{\downarrow\downarrow}\right\}$, $V_{\rm ddi}$ couples only $\ket{\uparrow\downarrow}$ to $\ket{\downarrow\uparrow}$, on the one hand, and $\ket{\uparrow\uparrow}$ together with $\ket{\downarrow\downarrow}$ on the other hand. In the latter case, the two pair states are separated in energy by several tens of GHz, and thus the effect of the dipolar coupling is negligible. In contrast, in the first case, the two pair states are always degenerate, and thus the dipolar coupling is resonant. Moreover, for those pair states, we always have $\Delta M=0$ (since the two atoms just exchange their states), so only the first term in (\textcolor{blue}{S}\ref{Eq:VdipSph}) survives. Therefore, when restricted to the two pair states $\ket{\uparrow\downarrow},\ket{\downarrow\uparrow}$, the interaction Hamiltonian takes the simple form:
\begin{equation}
\label{Eq:VdipZeroDeg}
\begin{aligned}
V_{\rm{ddi}} = \frac{1}{4 \pi \epsilon_0} \frac{1-3\cos^2\theta}{2R^3} \left(  d_{1,+} d_{2,-} + d_{1,-} d_{2,+} + 2d_{1,0}d_{2,0} \right)
\end{aligned}
\end{equation}
or, in a matrix form in the basis $\big\{\ket{\uparrow\downarrow},\ket{\downarrow\uparrow}\big\}$: 
\begin{equation}
V_{\rm{ddi}} = \frac{1}{4 \pi \epsilon_0} \frac{1-3\cos^2\theta}{R^3} 
\left( 
\begin{array}{cc}
0&\tilde{C_3}\\
\tilde{C_3}&0\\ 
\end{array}
\right).
\end{equation}
There, $\tilde{C_3}\equiv\bra{\uparrow\downarrow} d_{1,+} d_{2,-} + d_{1,-} d_{2,+} + 2d_{1,0}d_{2,0}\ket{\downarrow\uparrow}/2$. Note that in this matrix element, only one of the three terms of the sum contributes, depending on the $m_j$ values of the $\ket{\uparrow}$ and $\ket{\downarrow}$ states ~\cite{NoteS1}.

Now, in the context of spin Hamiltonians, it is desirable to rewrite the pseudo-spin operators in terms of the Pauli matrices. A first step towards this is then to denote the single-atom operators $\ket{\uparrow}\bra{\downarrow}$ as $\sigma_{\uparrow\downarrow}$, and we end up with 
\begin{equation}
V_{\rm{ddi}} =  \frac{C_3(\theta)}{R^3} \left( \sigma_{\uparrow\downarrow}^{(1)}\sigma_{\downarrow\uparrow}^{(2)}+{\rm \;h.\,c.} \right) 
\end{equation}
where $C_3(\theta)=\tilde{C_3}(1-3\cos^2\theta)$, and ${\rm h.\,c.}$ stands for hermitian conjugate. In Eq. (1) of the main text, to conform to the standard notations of the ladder operators $\sigma_\pm=\sigma_x\pm i\sigma_y$ defined from the Pauli matrices in the context of spin systems, we have written $\sigma_+$ (resp. $\sigma_-$) for $\sigma_{\downarrow\uparrow}$  (resp. $\sigma_{\uparrow\downarrow}$). We emphasize here that the $\pm$ subscripts in this latter notation refer to the pseudo-spin space, and have nothing to do with the $\pm$ subscripts entering the definition of the spherical dipole (\textcolor{blue}{S}\ref{Eq:DipoleOperatorsDefinition}), which are related to the orientation of the dipoles with respect to the quantization axis in real space ~\cite{NoteS2}.

\section{Simulation of the spin excitation dynamics in a chain}

To investigate the origin of the reduced contrast and damping of the oscillations in our experiment, we solve the optical Bloch equations (OBEs) for the system of three atoms in the basis spanned by the states $\left\{ \ket{g}, \ket{\uparrow}, \ket{\downarrow} \right\}$. By including the ground state $\ket{g}$ we can account for imperfections in the preparation of the initial configuration due to both spontaneous emission through the intermediate state, and the presence of the always resonant dipole-dipole interaction between the states $\ket{\uparrow \downarrow}$ and $\ket{\downarrow \uparrow}$. We simulate the full experimental sequence, including the optical and microwave pulses to the Rydberg states, the time evolution under the \textit{XY} Hamiltonian [(1) of the main text], and the final state readout. 
The total Hamiltonian of the system
\begin{equation}
\label{Eq:HtotMW}
\begin{aligned}
H_{\rm{tot}} =  H_{\mathit{A\mbox{-}L}} + H_{\rm{int}} \ ,
\end{aligned}
\end{equation}
is composed of two terms describing, respectively, the coupling of the atoms to the light and microwave fields, and the dipole-dipole interaction. The first term, $H_{\mathit{A\mbox{-}L}}$, reads: 
\begin{equation}
\label{Eq:HLMW}
\begin{aligned}
H_{\mathit{A\mbox{-}L}} = \hbar \sum_{i} \frac{\Omega_{\rm{L}}^{(i)}}{2} \left(\sigma_{\uparrow g}^{(i)} + \sigma_{g \uparrow}^{(i)}\right) + \frac{\Omega_{\rm{MW}}}{2} \left(\sigma_{\downarrow \uparrow}^{(i)} + \sigma_{\uparrow \downarrow}^{(i)}\right) - \delta_{\rm{L}}^{(i)}\left(\sigma_{\uparrow \uparrow}^{(i)} + \sigma_{\downarrow \downarrow}^{(i)}\right) \ ,
\end{aligned}
\end{equation}
where $\sigma_{nn'}^{(i)}=\ket{n^{(i)}}\bra{n'^{(i)}} (n, n' \in \left\{g, \uparrow, \downarrow \right\})$ are the transition and projector operators for atom at site $i$, and the parameters $\Omega_{\rm{L}}^{(i)}$, and $\delta_{\rm{L}}^{(i)}$ are the optical Rabi frequencies, and laser detunings for the transition $\ket{g} \leftrightarrow \ket{\uparrow}$ respectively. These parameters are position-dependent to account for the slightly different Rabi frequencies and differential light shifts experienced by the atoms due to the gaussian profile of the excitation laser beams. For the microwave Rabi frequency $\Omega_{\rm{MW}}$ for the transition $\ket{d} \leftrightarrow \ket{p}$, we neglect any inhomogeneity of the microwave field at the local position of the atoms. Finally, the interaction term is given by the Hamiltonian:
\begin{equation}
\label{Eq:HintMW}
\begin{aligned}
H_{\rm{int}} = \frac{1}{2}\sum_{i\neq j} \frac{C_3}{R_{i j}(t)^3} (\sigma_{\downarrow \uparrow}^{(i)} \sigma_{\uparrow \downarrow}^{(j)} + \sigma_{\uparrow \downarrow}^{(i)} \sigma_{\downarrow \uparrow}^{(j)}) \ .
\end{aligned}
\end{equation}
To account for the thermal motion of the atoms during the experiment we introduce a time dependency in the interatomic distance $R_{i j}(t)=|({\bs r}_i^0+{\bs v}_i^0 t)-({\bs r}_j^0+{\bs v}_j^0 t)|$.

The Lindblad operator includes a sum over the decay channels for each atom, 
\begin{equation}
L[\rho]=\frac{1}{2}\sum_i{(\gamma_i+\gamma_\uparrow)\left(2\sigma_{g \uparrow}^{(i)} \rho \sigma_{\uparrow g}^{(i)}-\sigma_{\uparrow \uparrow}^{(i)} \rho - \rho \sigma_{\uparrow \uparrow}^{(i)}\right)} + \gamma_\downarrow \left(2\sigma_{g \downarrow}^{(i)} \rho \sigma_{\downarrow g}^{(i)}-\sigma_{\downarrow \downarrow}^{(i)} \rho - \rho \sigma_{\downarrow \downarrow}^{(i)}\right) \ .
\end{equation}
where $\gamma_i$ are the effective damping rates, and  $1/ \gamma_\uparrow$ and $1/ \gamma_\downarrow$ are the lifetimes for the $\ket{\uparrow}$ and $\ket{\downarrow}$ Rydberg states. The dampings $\gamma_i$ (with $1/\gamma_i\sim 1\,\mu{\rm s}$) are mainly due to off-resonant spontaneous emission through the intermediate state $\ket{5P_{1/2}}$ and are only present during the optical pulses. For the lifetimes of the Rydberg states we use the effective values  ($1/ \gamma_\uparrow\sim 101 \mu$s ; $1/ \gamma_\downarrow\sim 135\mu$s) calculated by Beterov \textit{et al.} [Ref. 32 of the main text], as a single decay channel to the ground state $\ket{g}$. All the parameters, i.e., the single-atom Rabi frequencies $\Omega_i$, laser detunings $\delta_i$, and dampings $\gamma_i$ are measured independently by recording single-atom Rabi oscillations on each site of the array. 

The final result of the simulation is obtained by averaging the solution of the OBEs over 100 realizations, starting from random atom positions ${\bs r}_i^0$, ${\bs r}_j^0$, with rms value $\sigma_r = \sqrt{k_B T/ m \omega_\bot ^2}\simeq 120$ nm (where $k_B$ is the Boltzmann constant and $\omega_\bot \sim 90$ kHz is the measured radial frequency of the atoms of mass $m$ in the trap), and velocities ${\bs v}_i^0$, ${\bs v}_j^0$ (rms $\sigma_v = \sqrt{k_B T/ m}\simeq 70$ nm/$\mu$s).

\section{Readout detection errors}

In our experiment, the outcome of every realization is a binary state: either the atom is recaptured or it has escaped the trapping region at the end of the sequence. If the atom is recaptured we assume it is in the ground state. On the contrary, a loss in interpreted as the atom being in a Rydberg state ($\ket{\uparrow}$ or $\ket{\downarrow}$), since Rydberg atoms are not trapped in our tweezers. However, due to the finite temperature ($\simeq 50~\rm{\mu}$K) of the atoms and collisions with the background gas, any atom has a probability $\varepsilon = \varepsilon(t)$ to be lost during a sequence of total length $t$, independently of its internal state [Ref 20 of the main text]. Therefore, there exists a probability to falsely detect the state of the atom, which increases with the length of the sequence. For a single atom, the recapture probability $P_1$ is therefore $P_1= (1-\varepsilon)P_g$, where $P_g$ is the actual probability for the atom to be in the ground state. In the case of a loss, $P_0$ is related to the actual Rydberg state populations $P_\uparrow$ and $P_\downarrow$ by $P_0=(1-\varepsilon)(P_\uparrow+P_\downarrow) + \varepsilon$. As an example, in the case of three atoms the relation between the observed recapture probability $P_{100}$ and the actual state populations $P_{j_1 j_2 j_3}$ ($j_k \in \left\{g, \uparrow, \downarrow \right\}$) reads:
\begin{align}
P_{100} &= (1-\varepsilon) [P_{g\uparrow\uparrow}+P_{g\uparrow\downarrow}+P_{g\downarrow\uparrow}+P_{g\downarrow\downarrow}+ \varepsilon (P_{g \uparrow g}+P_{g \downarrow g}+ P_{gg\uparrow}+P_{gg\downarrow}) + \varepsilon^2 P_{ggg}]
\end{align}

To measure $\varepsilon$, we perform an experiment where we trap three ground state atoms in the line, switch the trap off for a variable duration $t$, and measure the populations at the end of the sequence. 

\begin{figure*}[t!]
\begin{center}
\includegraphics[width=18cm]{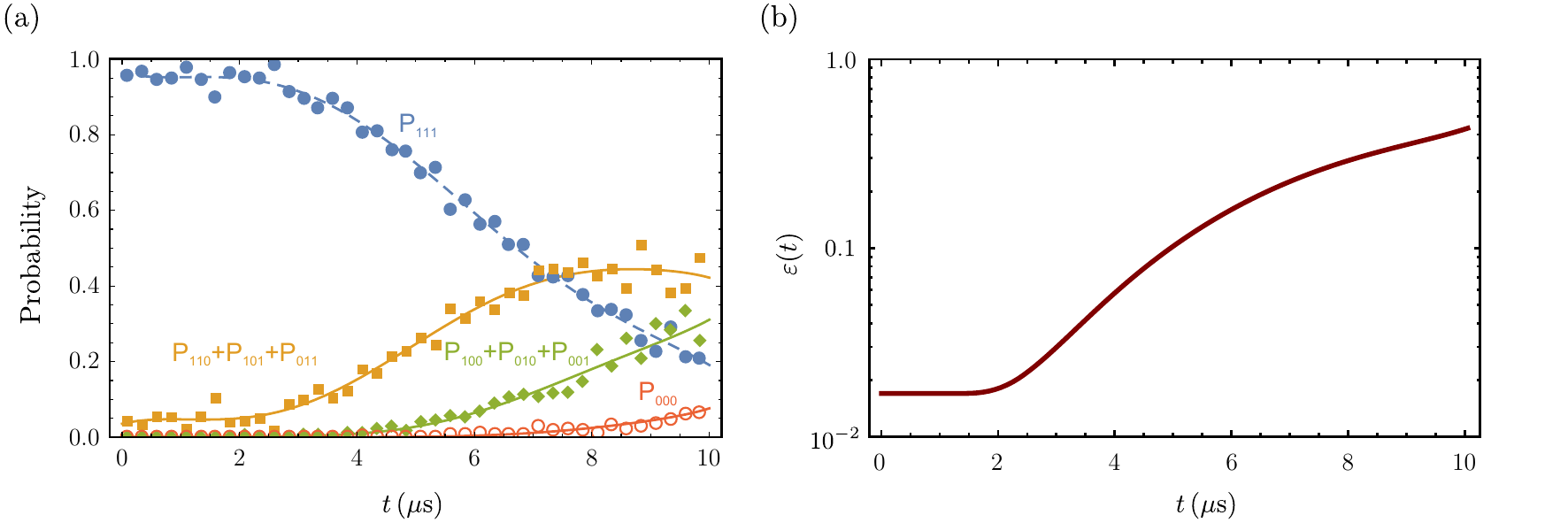}
\end{center}
\caption{(a) Recapture probabilities $P_{111}$ (solid circles), $P_{110}$+$P_{101}$+$P_{011}$ (square symbols), $P_{100}$+$P_{010}$+$P_{001}$ (diamonds), and $P_{111}$ (open circles) for three atoms in a line as a function of the release trapping time $t$. The dashed line is the polynomial fit to the data from which we extract $\varepsilon$. Solid lines are the prediction of our loss model with the measured $\varepsilon$, which is shown in logarithmic scale in (b). The nonzero value of $\varepsilon(t)$ for $t\to 0$ is due to losses induced by background-gas collisions during the $\sim 200$~ms duration of a full experimental sequence.}
\label{fig:figS1}
\end{figure*}

We extract $\varepsilon$ from a polynomial fit of the measured $P_{111}$, where $P_{111} = (1-\varepsilon)^3$. We then evaluate the evolution of the rest of the recapture probabilities:
\begin{equation}
\label{Eq:EpsT}
\begin{aligned}
& P_{011} + P_{101} + P_{110} = 3\,\varepsilon (1-\varepsilon)^2 \ , \\
& P_{100} + P_{010} + P_{001} = 3\,\varepsilon^2(1-\varepsilon) \ , \\
& P_{000} = \varepsilon^3 \ .
\end{aligned}
\end{equation}
The comparison of the calculated recapture probabilities (solid lines in Fig. S1) with the experimental test supports the consistency of our loss model.

\section{Origin of the finite contrast of the oscillations}

In Fig. 2(b) of the main text, the measured contrast of the oscillations in the populations  $P_{01}$ and $P_{10}$ is around $60\%$. This reduced contrast arises from two effects. The first one is imperfections in the preparation of the state $\ket{\uparrow \downarrow}$  mainly due to (i) inefficient optical pumping of the atoms in the state $\ket{g}$ and (ii) spontaneous emission from the intermediate state $\ket{5P_{1/2},F=2,m_F=2}$ to the states $\ket{5S_{1/2},F=1,m_F=1}$, $\ket{5S_{1/2},F=2,m_F=1}$, and $\ket{5S_{1/2},F=2,m_F=2}$. The second effect is that during the second and third optical pulses, of typical duration 100~ns, one cannot neglect totally the effect of the dipolar interactions between states $\ket{\uparrow \downarrow}$ and $\ket{\downarrow\uparrow }$: the interaction-induced energy splitting between the two states is then 0.6~MHz, yielding a phase accumulation of $\sim0.2$ rad, which is not totally negligible. The same effect also occurs for three atoms, as described in the main text. The full simulation (following the approach described in sections S.2 and S.3 above) of the two-atom system reproduces well the observed data.

\section{Full dataset for the three-atom spin chain dynamics}

The evolution of the system of three spins aligned along a chain coupled by long range dipole-dipole interactions is displayed in Fig. S2. Here we show, additionally, the probabilities not shown for clarity in Fig. 3 of the main text.

Solid lines are the result of the simulation through the OBEs as described in section S.1., where we also include the loss-error correction (section S.2.). The atom loss effect is clearly visible in the probability $P_{000}$ at long interaction times. This effect further reduces the contrast of the oscillations for the populations $P_{100}$, $P_{010}$, and $P_{001}$. The overall agreement of the simulations is very good, supporting the possibility to extend the experiments to larger two dimensional systems.
\begin{figure*}[h!]
\centering
\includegraphics[width=18cm]{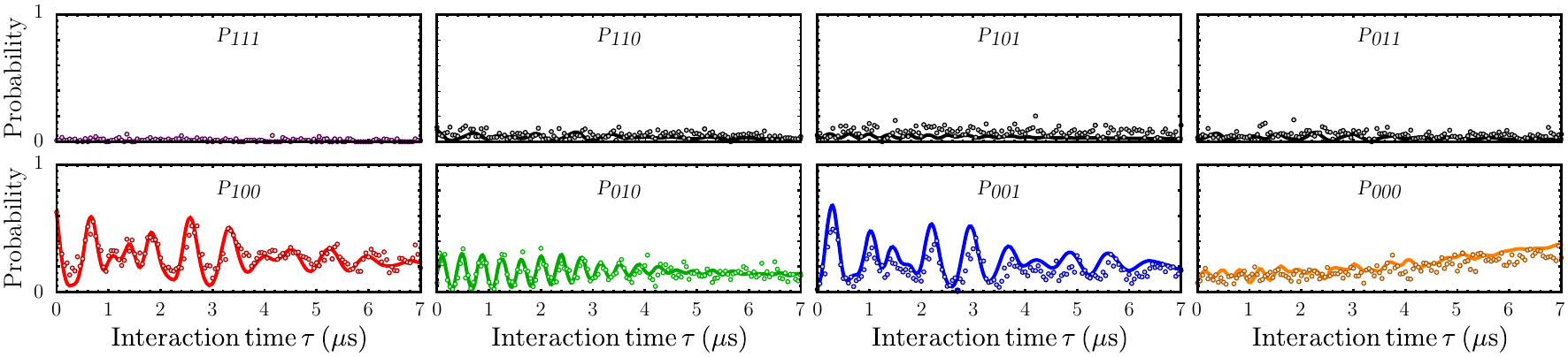}
\caption{Spin excitation dynamics along a chain of three Rydberg atoms from a system initially prepared in $\ket{\uparrow \downarrow \downarrow}$. The complete set of experimental probabilities (symbols) and the result of the model (solid lines), including temperature effects and atom losses are shown.}
\label{fig:figS2}
\end{figure*}

\section{Effect of the atom temperature in the dynamics}

In an attempt to investigate the limitations of the finite temperature of the atoms in the observed spin dynamics, especially in view of experiments with larger number of atoms, we simulate the evolution of the system at $T=10\, \mu$K. The temperature of the atoms enters the simulation in (i) the rms extension of the random atomic positions and velocities in the time-dependent dipolar couplings, and (ii) the atom recapture probability $\varepsilon(t,T)$.  To include (i) we perform Monte-Carlo simulations of the trajectories of the single atoms, taking into account the expected energy distribution of an atom in the trap ~\cite{TuchendlerSup2008}.

The result is shown in Fig. S3, where we compare the simulation for $T = 10 \,\mu$K (solid lines) with the curves obtained at $T = 50\, \mu$K (dashed lines). In this case the damping of the oscillations are drastically reduced for times longer than $\sim 4\,\mu$s. This suggests that a reduction of the temperature of the atoms by only one order of magnitude would be enough to make motional effects negligible for the timescales used in the experiment, so as to enable the nearly ideal quantum simulation of larger spin systems.
\begin{figure*}[h!]
\centering
\includegraphics[width=18cm]{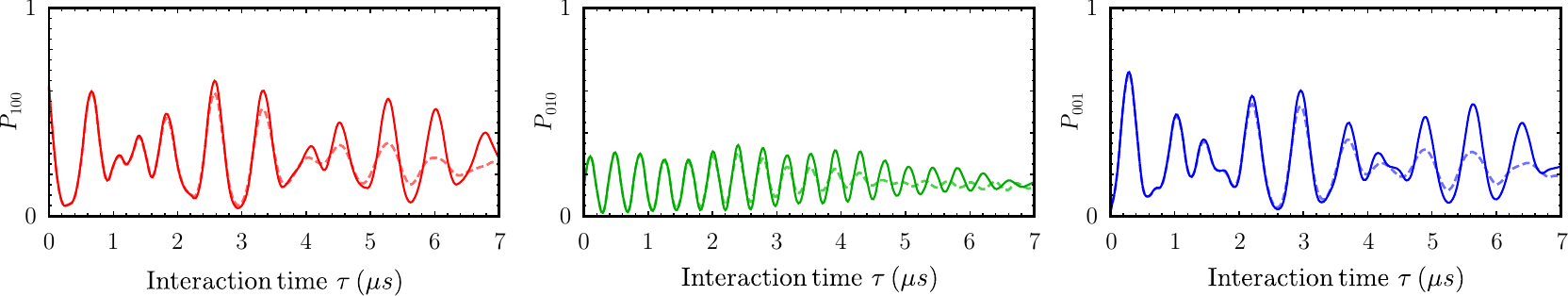}
\caption{Simulated recapture probabilities $P_{100}$, $P_{010}$, and $P_{001}$ for a single atom temperature of $T = 10\, \mu$K  (solid lines), and for $T = 50\,\mu$K (dashed lines).}
\label{fig:figS3}
\end{figure*}

\section{Extension of the system to longer spin chains}

Considering the temperature of the atoms as the main source of dephasing in the excitation dynamics, we now explore its effect for a larger spin chain. We simulate the evolution of a system of $N=20$ atoms in a line, separated by $20\,\mu$m under the Hamiltonian (1) (main text). In Fig. S4 we show the results for two different atom temperatures, $T=0\, \mu$K (a), and for $T=50\, \mu$K (b).  The simulation assumes perfect preparation of the initial state $\ket{1\uparrow}=\ket{\uparrow\downarrow...\downarrow}$ (imperfect preparation would reduce the overall amplitude of the oscillations, but would not induce any extra loss of coherence) and accounts for temperature effects through both (i) time-dependent distances $R(t)$ appearing in the dipolar couplings and (ii) the finite recapture probability $\varepsilon$ (as discussed in the previous sections). To account for (ii) we scale the obtained excitation probability by the factor $[1-\varepsilon(t)]^{N-1}$. 

Dephasing in the dynamics is appreciable for interaction times longer than  $\sim 4\,\mu$s, but still allows for an unambiguous observation of the spin dynamics.  This means that even with the finite temperature of the atoms in our tweezers, motional effects should not prevent the observation of coherent spin exchange in larger systems. Therefore, as stated in the conclusion of the main text, it is the stochastic character of the atom loading into the traps which is currently the main limitation for scalability, and we plan to explore several quasi-deterministic loading schemes that have already been demonstrated in other groups (refs. [37-40] of main text).

\begin{figure*}[h!]
\centering
\includegraphics[width=18cm]{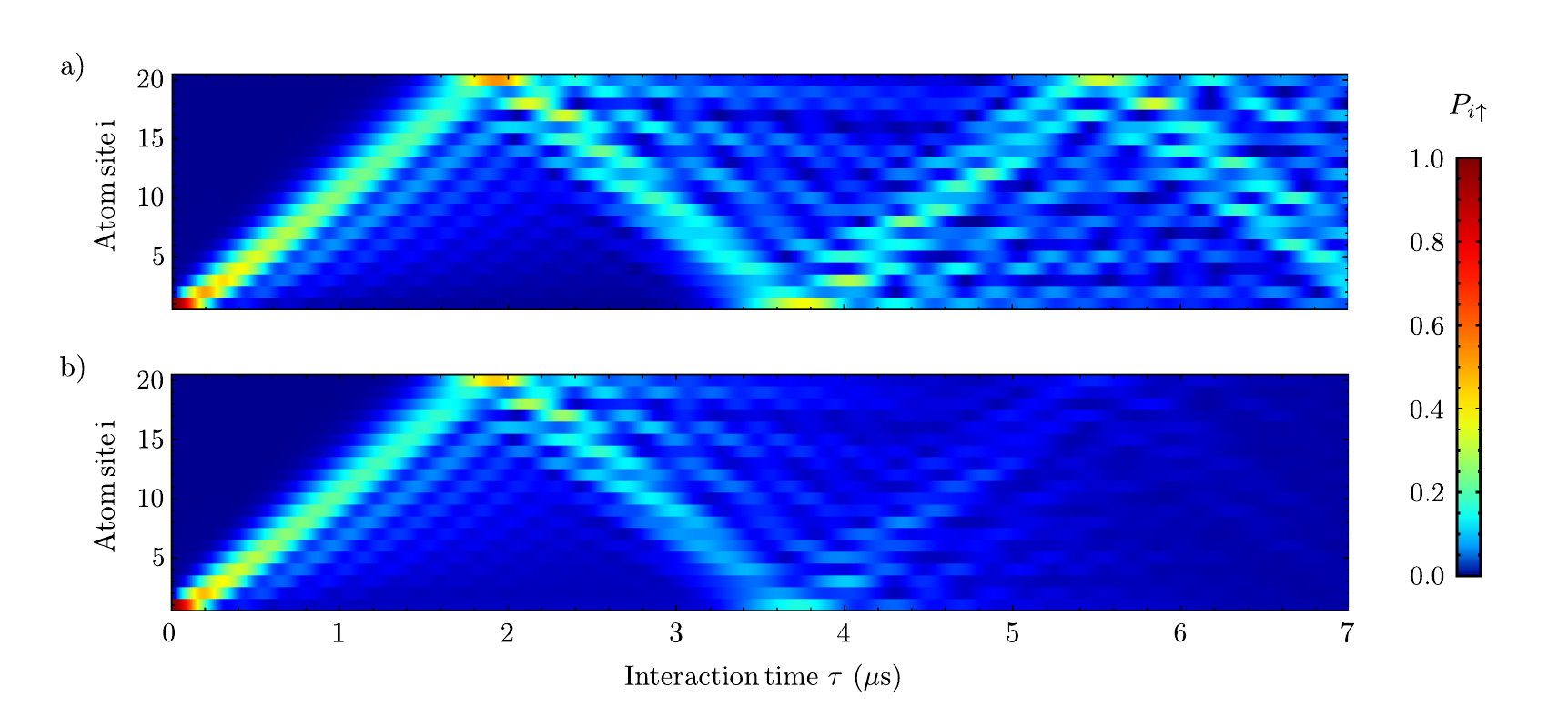}
\caption{Simulated probabilities $P_{i\uparrow}$ of finding atom $i$ in the state $\ket{\uparrow}$ after an interaction time $\tau$ for single atoms with a temperature $T = 0\, \mu$K  (a), and $T = 50\,\mu$K (b). The system is initially prepared in the state $P_{1\uparrow}$. }
\label{fig:figS4}
\end{figure*}


\begin{thebibliography}{9}
\bibitem{NoteS1}
{For instance, for the states used in the main text, $\ket{\uparrow}=\ket{62D_{3/2},m_j=3/2}$ and $\ket{\downarrow}=\ket{62P_{1/2},m_j=1/2}$, only the term $d_{1,+}d_{2,-}$ has a nonzero contribution. If one used for instance the states $\ket{\uparrow}=\ket{nD_{3/2},m_j=3/2}$ and $\ket{\downarrow}=\ket{n'P_{3/2},m_j=3/2}$, then only the term $2d_{1,0}d_{2,0}$ would contribute.}
\bibitem{NoteS2}
{One should not be confused by the notations, and believe that for instance the $\sigma_+\sigma_-$ term arises from the $d_+ d_-$ term of the dipole dipole interaction: the $\sigma_+\sigma_-$ term comes only from the fact that the resonant dipole-dipole interaction exchanges the two spin states, and this can arise even from a $d_0d_0$ term. }
\end{thebibliography}
\end{document}